\documentstyle{l-aa}

\input psfig.tex
\ifx\undefined\AALPLAIN \else
\def\.#1{{\accent"C7 #1}}
\def\dot{\mathaccent"70C7 }
\fi

\begin{document}

\thesaurus{03(02.01.2; 02.02.1; 11.01.2; 02.18.7)}

\title{Vertical Structure and Spectrum of Accretion Disks in 
Active Galactic Nuclei}

\author{T. D\" orrer\inst{1}
\and    H. Riffert\inst{2}
\and    R. Staubert\inst{1}
\and    H. Ruder\inst{2}
}

\offprints{R. Staubert}

\institute{Institut f\"ur Astronomie und Astrophysik, Astronomie,
Universit\"at T\"ubingen, Waldh\"auserstr. 64, D-72076 T\"ubingen, Germany
\and
Institut f\"ur Astronomie und Astrophysik, Theoretische Astrophysik,
Universit\"at T\"ubingen, Auf der Morgenstelle 10, D-72076 T\"ubingen, Germany
}

\date{Received date; accepted date}

\maketitle

\begin{abstract}
Radiation emitted from an accretion disk around a massive black hole
is a widely discussed model for the UV/soft X--ray excess emission
observed in the spectra of many AGN. 
A self--consistent calculation of the
structure and the emerging spectrum of geometrically thin $\alpha $--accretion
disks in AGN is presented.
The central object is assumed to be a Kerr black hole; full relativistic
corrections are included. The local dissipation of gravitational
energy is assumed
to be entirely due to turbulence.
Since these disks are mainly supported by radiation pressure, 
the assumption that
the viscous energy production is proportional to the total pressure
leads to diverging temperature structures in the upper parts of the disk,
where the total optical depths is small. We therefore modify the standard
expression for the turbulent viscosity by taking into account the radiative
energy loss of turbulent elements in an optically thin regime.
Compton scattering is treated in the
Fokker--Planck approximation using the Kompaneets oparator.
The absorption cross section contains only free--free
processes for a pure hydrogen atmosphere.
We present several calculations for various parameters such as
the accretion rate $\dot{M}$, the viscosity parameter $\alpha$, the
specific angular momentum $a$ of the black hole and the inclination angle
$\Theta _{0}$ of the observer.
The resulting temperature and density profiles show that the disks are
optically thick for Compton scattering and effectively
optically thin for most frequencies.
This leads to spectra that are
diluted with respect to the photon number but contain a Comptonized
high energy tail in order to carry the required energy flux.
In addition the electron temperature deviates strongly from the equilibrium
temperature. For a model with $M=10^{8}M_{\odot}$, $\dot{M}=0.3$,
$\alpha=1/3$ and $a/M=0.998$ the fraction of total flux
emitted 
in the soft and hard X--ray band ($>2.4\cdot 10^{16}Hz$)
for $\Theta _{0} = 0\degr ,\, 41\degr ,\, 60\degr ,\, 70\degr $ 
and $90\degr$ is 
$36\%$, $50\%$, $64\%$, $78\%$ and $93\%$,
respectively.
Therefore, the model can in general account for the observed soft X--ray 
excess. 

\end{abstract}

\section{Introduction}
Dissipation of gravitational energy by accretion onto a massive black hole is 
regarded as the
origin of the enormous luminosities of active galactic nuclei (AGN).   
Since the accreted matter is likely to carry some angular momentum,
it is common to invoke the presence of an accretion disk.
A strong observational evidence for such disks comes from 
the spectral flattening
in the optical/UV seen in many AGN, the so--called ''big blue bump''
(e.\ g.\ Shields \cite{shields}; Malkan \& Sargent \cite{malkan}).
This spectral feature is thought to be due to thermal emission from
an accretion disk and perhaps extends into the soft X--ray band,
producing a steep excess emission known as the ''soft excess''
(e.\ g.\ Arnaud et al.\ \cite{arnaud}; Wilkes \& Elvis \cite{wilkes}; 
Turner \& Pounds \cite{turner}). 
Other evidence comes from the reflection hump and iron emission line
observed in AGN (e.\ g.\ Pounds et al.\ \cite{pounds}).

The standard theory of thin accretion disks is largely based on
the fundamental papers of  
Shakura \& Sunyaev (\cite{shakura}, hereafter SS73) and Novikov \& Thorne 
(\cite{novikov}, hereafter NT73).
The simplest way to calculate the emitted disk spectrum is to assume
that the disk is geometrically thin but optically thick and radiates 
locally as a blackbody. The effective temperature $T_{\rm eff}$
of the blackbody is then solely determined by the dissipated flux per
unit area.
These simple disk models, however, are not able to produce soft X--rays
as long as they do not become super--Eddington (e.\ g.\ Bechtold et al.\ 
\cite{bechtold}).
Real disk spectra will differ from the blackbody approximation:
for sufficiently high accretion rates and viscosity parameters, 
the accretion disk eventually becomes 
effectively optically thin at small radii and the gas temperature strongly
deviates from the equilibrium temperature in such cases.
Even in the optically thick 
case the local spectra differ from
the blackbody, because 
the scattering opacity 
dominates over absorption in the soft X--ray regime and 
there exists a temperature gradient 
in the vertical direction.

In the last few years several calculations of increasing accuracy
have been performed to determine the structure and emission spectrum 
of accretion disks around massive black holes. 
Contrary to the standard Newtonian disk model, NT73 and Page \& Thorne
(\cite{page}) calculated the
effects of general relativity on the disk structure.
The propagation of photons from the disk surface to
a distant observer was treated by Cunningham (\cite{cunningham}).
On the other hand, SS73, Czerny \& Elvis (\cite{czerny}), and 
Wandel \& Petrosian (\cite{wandel})
concentrated on a proper calculation of the radiative transfer 
by discussing the effects of Comptonization in a simple analytic
manner. Laor \& Netzer (\cite{laor89}) and Laor et al.\ (\cite{laor90}) 
included
relativistic effects as well as free--free and bound--free opacities in their
numerical computations.  
Most calculations of model spectra to date, however, made use of 
an averaging over the vertical direction.  
Ross, Fabian \& Mineshige (\cite{ross}, hereafter RFM) have calculated 
the vertical temperature
profile and atomic level populations 
in the radiation pressure dominated inner regions of the disk 
for a given constant vertical density profile, using the 
Kompaneets equation to treat Compton scattering.     
Relativistic corrections for a non--rotating black hole 
were incorporated in this code by Yamada et al.\ (\cite{yamada}).
Shimura \& Takahara (\cite{shimura}) and
Shimura \& Takahara (\cite{shimura2})
have calculated the vertical structure and
radiation field of a Newtonian disk self--consistently,
using the ad hoc assumption that the local
viscous heating rate is proportional to the mass density $\rho$.

Here we have also calculated the z--structure simultaneously with the
radiation field of the disk.
But we use a different viscosity description. We assume that the
local energy production is caused by turbulence. The standard
$\alpha $--description (viscosity proportional to the total pressure)
leads to diverging temperature profiles in the upper optically thin
regions of the disk,
because viscous heating always exceeds
radiative cooling. We therefore propose a modification of the standard
$\alpha $--model, which includes the radiative cooling of the
turbulent elements.
The frequency dependent radiative transfer equation
is solved in the Eddington approximation
and the effects of Compton scattering are treated by the Kompaneets
equation (Kompaneets \cite{kompanaets}).
Relativistic corrections on the local disk structure
are introduced according to Riffert \& Herold (\cite{riffert})
for rotating and non--rotating black holes.
In this paper we present solutions of the local structure
and the emission spectrum of accretion disks for different 
input parameters (accretion rate, viscosity parameter).
Integrated disk spectra, as seen from a distant observer, are calculated
by the use of a transfer function (e.\ g.\ Cunningham \cite{cunningham};
Speith et al.\ \cite{speith}).
In Sect.\ 2 the basic equations and viscosity description of our model
are formulated. In Sect.\ 3 numerical results are presented.
Finally, our results and conclusions are summarized in Sect.\ 4.

\section{The Model}

\subsection{Basic Equations}

We adopt the standard geometrically thin $\alpha$--accretion disk model,
i.e. the disk height $H$ is much smaller than the radius $r$,
and the dominant velocity is given by the Keplerian motion around
the central mass. The disk is assumed to be in a stationary and
rotationally symmetric state, thus all functions depend only on the
radial and vertical coordinates $r$ and $z$.
The relativistic disk structure has been calculated by NT73 and
subsequently by Riffert \& Herold (\cite{riffert}), 
correcting a term in the vertical
pressure balance. According to this paper we define four relativistic 
correction factors with respect to the standard Newtonian
disk model depending on the mass $M$ and the specific angular 
momentum $a/M$ of the central black hole:

\begin{eqnarray}
A&=& 1-\frac{2GM}{c^2 r} +\frac{a^2}{c^2 r^2}                \nonumber \\
B&=& 1-\frac{3GM}{c^2 r} +\frac{2a\sqrt{GM}}{c^2 r^{3/2}}    \nonumber \\
C&=& 1-\frac{4a\sqrt{GM}}{c^2 r^{3/2}} +\frac{3a^2}{c^2 r^2} \nonumber \\
D&=& \frac{1}{2\sqrt{r}} \int_{r_i}^r
            \frac{x^2 c^2 - 6xGM + 8a\sqrt{xGM} - 3a^2}
                 {\sqrt{x}\left( x^2 c^2 - 3xGM + 2a\sqrt{xGM} \right)}\ dx~~.
\end{eqnarray}
$r_{i}$ is the inner disk radius located at the position of the 
last stable circular orbit;
the gravitational constant and the velocity of light are denoted 
by $G$ and $c$. 

The hydrostatic equilibrium in vertical direction
for a thin accretion disk is given by

\begin{equation}
\label{hyd}
\frac{dP}{dz} = -\rho g_{z} = -\rho \frac{GM}{r^{3}} z \frac{C}{B} \, ,
\end{equation}
where $P$, $\rho $, and $g_{z}$ are the total pressure, 
the mass density, and the gravitational acceleration, respectively. 

The radiative transfer is solved in the Eddington approximation,
and Compton scattering is treated in the Fokker--Planck
approximation using the Kompaneets operator.
The absorption cross section contains only free--free processes
for a pure hydrogen atmosphere.
Induced contributions to the radiative processes have been neglected throughout.

Assuming the plasma to be in a state of local thermodynamic equilibrium
(LTE) the first two moments of the radiative transfer
equation for the spectral energy density $U_{\nu}$ and the spectral
flux $F_\nu$ read

\begin{equation}
\label{rad1}
\frac{\partial F_{\nu}}{\partial z}
= \kappa _{\nu}^{\rm ff}\rho c [W_{\nu}-U_{\nu}] 
+\kappa _{T} \rho \frac{8\pi h^{2}}{m_{e}c^{4}}\nu
\frac{\partial}{\partial \nu} \nu ^{4}
\left[ n_{\nu}
+\frac{kT}{h}\frac{\partial n_{\nu}}{\partial \nu}\right]
\end{equation}
\begin{equation}
\label{rad2}
\frac{1}{3} \frac{\partial U_{\nu}}{\partial z}=
-\frac{\rho}{c}[\kappa _{T} +\kappa _{\nu}^{\rm ff}] F_{\nu} \, .
\end{equation}
Here $W_{\nu} = 8\pi h \nu ^{3}c^{-3}\exp(-h\nu / kT)$ 
is the Wien function which serves as the equilibrium spectrum in this case,
$\kappa_{\nu}^{\rm ff}$ is the free--free opacity, and $\kappa_{T}=
0.4~{\rm g/cm^2}$ is the Thomson opacity.
The photon occupation number $n_{\nu}$ can be expressed in terms of 
$U_{\nu}$ 

\begin{equation}
\label{phot}
n_{\nu}=\frac{c^{3}}{8\pi h \nu ^{3}}U_{\nu} \, .
\end{equation}
The free--free opacity is given by

\begin{equation}
\label{opaz}
\kappa _{\nu}^{\rm ff}=1.3\cdot 10^{56}\rho T^{-1/2}
\nu ^{-3} \bar{g}_{\rm ff}\: cm^{2}\, g^{-1}\, ,
\end{equation}
where $T$ is the gas temperature, $k$ is the Boltzmann constant,
and $\bar{g}_{\rm ff}$ is the thermal average free--free Gaunt factor 
(Karzas \& Latter \cite{karzas}; Carson \cite{carson}).

In the Eddington approximation the equation of state can be written as

\begin{equation}
\label{state}
P= P_{gas}+P_{rad}= \frac{k}{\mu m_{u}}\rho T + \frac{1}{3}U \, ,
\end{equation}
where $U=\int _{0}^{\infty}U_{\nu}\, d\nu$ is the total energy density
of the radiation field,
$m_{u}$ is the atomic mass unit and $\mu$ is the mean molecular weight  
with $\mu =1/2$ for a fully ionized hydrogen atmosphere.
  
For a Keplerian disk, the viscous heating rate per unit 
volume $\dot{Q} _{visc}$ is related to the
$\phi$--$r$--component of the viscous stress tensor ${\bf t}$ by

\begin{equation}
\label{visc}
\dot{Q} _{visc}= \frac{3}{2}\sqrt{\frac{GM}{r^{3}}}
t_{\phi r}\frac{A}{B} \, ,
\end{equation}  
and $t_{\phi r}$ can be expressed in terms of velocity gradients
leading to the result

\begin{equation}
\label{shear}
t_{\phi r} = \frac{3}{2} \eta \sqrt{ \frac{GM}{r^3} } \frac{A}{B}~,
\end{equation}
with the shear viscosity $\eta$.

In a steady state this heating rate must be balanced by the total
radiative cooling, since convection is not considered in this model.
Integration of the right--hand side
of equation (\ref{rad1}) over frequency $\nu$
leads to the following energy balance equation 

\begin{equation}
\label{ebal}
\dot{Q} _{visc}=\frac{dF}{dz}=\int_{0}^{\infty} 
\frac{\partial F_{\nu}}{\partial z} \, d\nu
\, .
\end{equation}
The disk model is so far described by the equations
(\ref{hyd}), (\ref{rad1}), (\ref{rad2}), (\ref{state}) and (\ref{ebal})
which can be solved numerically when proper
boundary conditions are imposed. The only unspecified process is
the source of viscosity and this will be discussed in the next section.

Because of symmetry reasons the inner boundary condition in the
midplane of the disk ($z=0$) is taken as

\begin{equation}
\label{bound1}
F_{\nu}(z=0)=0 \, ~,
\end{equation}
and at the upper boundary (at $z=H$) we assume that the disk radiates
isotropically into the vacuum

\begin{equation}
\label{bound3}
F_{\nu}=\frac{c}{2}U_{\nu} \, .
\end{equation}
For a geometrically thin accretion disk the energy flux from
the disk surface ($z=H$) is

\begin{equation}
\label{bound2}
F(z=H)= \int_{0}^{\infty} F_{\nu} \, d\nu =
\frac{3GM\dot{M}}{8\pi r^{3}}\frac{D}{B} \, ,
\end{equation}
where $\dot{M}$ is the total mass flux through the disk.

Finally, the disk height $H$ is specified by the condition

\begin{equation}
\label{bound4}
\rho (z=H) =0
\end{equation}
We solve the above set of equations for the functions
$\rho$, $T$, $H$, $U_{\nu}$, and $F_{\nu}$ using a finite
difference scheme in both variables $z$ and $\nu$.
The vertical structure is resolved with 100 points on a
logarithmic grid, and 64 grid points are used in frequency space.
The resulting set of algebraic difference equations is then solved by
a Newton--Raphson method.

Once the vertical structure has been determined we can 
integrate the density profiles to obtain the 
surface density $\Sigma (r) = \int _{-H}^{+H} \rho (z) \, dz $
and the average radial velocity $|v_{r}| = \dot{M}/ \Sigma (r) $.
In order to justify the basic assumptions of the thin disk model
one must have $|v_{r}| \ll |v_{\phi}| $, where $v_{\phi}$ is the
Keplerian velocity.

\subsection{The Viscosity Description}

In the standard $\alpha$-model (SS73) the viscous stress tensor
component $t_{\phi r}$ is propotional to the pressure
\begin{equation}
\label{alpha}
t_{\phi r}= \alpha P~.
\end{equation}
This is a consequence of a number of assumptions: first, the
viscosity is caused by turbulence in the disk, and the magnitude
of $\eta$ can be estimated from a dimensional analysis
\begin{equation}
\label{eta}
\eta \approx \rho \, l_{turb} V_{turb}~,
\end{equation}
where $l_{turb}$ and $V_{turb}$ are the size and the typical
velocity of the largest turbulent eddies, respectively.
Next, the disk height $H$ is used for $l_{turb}$ which is obtained
from the hydrostatic equation (\ref{hyd})

\begin{equation}
H \approx \sqrt{ \frac{B\, r^3}{C\, GM} }
          \sqrt{ \frac{P_c}{\bar{\rho} } }~,
\end{equation}
where $P_{c}$ is the pressure in the midplane, and $\bar{\rho}$
stands for the average density.
The turbulent velocity $V_{turb}$ is assumed to be limited by the
local sound speed $c_s$. Otherwise shocks would develop and heat the
gas until the tubulence is subsonic. Approximating the ratio
$P_c / \bar{\rho}$ by the local value $P/\rho = c_s^2$ and inserting
the above relations into equation (\ref{shear}) for the viscous shear
leads to the expression (\ref{alpha}), where the constant $\alpha$
takes up all uncertainties in terms of numerical factors.
This entire procedure is quite successful for the standard disks
in cataclysmic variables which are dominated by the gas pressure.
If radiation pressure becomes important a number of difficulties
arise when the model is applied without modifications.
According to equation (\ref{rad1}) the radiative cooling
decreases in the upper layers of the disk where the density
drops exponentially. Since the radiation pressure remains
finite in this region, the total heating rate (\ref{visc})
cannot be balanced by radiative losses.
This could be cured by replacing $P$ in equation (\ref{alpha})
by the gas pressure, however, in an optically thick regime
the radiation behaves like a gas exchanging momentum with the
plasma and thus contributes to the viscous shear.
In addition, in a radiation dominated disk gravity is mainly balanced by
radiative forces and the disk height is then given by

\begin{equation}
\label{heightedd}
H \approx \frac{3 \dot{M} \kappa_F}{8\pi c} \frac{D}{C}~,
\end{equation}
where $\kappa_F$ is a flux weighted average opacity

\begin{equation}
\label{Rossel}
\frac{1}{\kappa _{F}}=\frac{\int_{0}^{\infty }\frac{1}{\kappa _{T} +
\kappa _{\rm ff}}
\frac{dU_{\nu}}{dz}\, d\nu }{ \int_{0}^{\infty } \frac{dU_{\nu}}{dz} \,
d\nu } \, .
\end{equation}
This height does not contain the sound speed.
In order to derive a consistent parameterization of the
turbulent viscosity for the radiation dominated case
we start from the dimensional analysis (\ref{eta}) using
again $l_{turb} = H$, where the disk height $H$ is
self--consistently calculated in our model. For the limitation of
the velocity $V_{turb}$ the same arguments as in SS73 are used,
i.e. $V_{turb}$ has to be small enough that no entropy due
to shock waves is generated in the flow. For a radiation dominated
shock in an optically thick medium

\begin{equation}
V_{turb} < \sqrt{\frac{P}{\rho}}~,
\end{equation}
but in an optically thin regime, where the photons are
able to escape freely
\begin{equation}
V_{turb} < \sqrt{\frac{P_{gas}}{\rho}}~.
\end{equation}
This follows from considering an isothermal shock in a
plasma with radiation when the Mach number approaches unity.
Assuming an isothermal shock structure is an approximation
for an optically thin environment.
Since this limiting value of $V_{turb}$ depends on the
optical depth we interpolate between the two extreme cases
\begin{equation}
\label{vturb}
V_{turb}= c_{s}\frac{ \tau +\sqrt{\beta} }{1+\tau }\, ,
\end{equation}
with the optical depth

\begin{equation}
\tau = \int_z^H \kappa_F \rho\, dz^\prime\, ,
\end{equation}
and the pressure ratio $\beta =P_{gas}/P$.
For the turbulent viscosity we now have
\begin{equation}
\eta = \alpha \rho H V_{turb}~,
\end{equation}
with a constant $\alpha$ which is a parameter in our disk model.

We will now consider the heating--cooling balance, integrate
equation (\ref{rad1}) over frequency, and use the relations (\ref{ebal})
and (\ref{visc})

\begin{eqnarray}
\frac{dF}{dz} &=& \bar{\kappa}_{\rm ff} \rho c \left[ a_{r}T^4 - U \right]
 + 4U\, \frac{\kappa_T \rho}{m_e c} \left[ kT - <\!\nu\!> \right] \nonumber \\
 &=& \frac{9GM}{4r^3} \frac{A^2}{B^2} \alpha \rho H V_{turb}~,
\end{eqnarray}
where $\bar{\kappa}_{\rm ff}$ is some appropriately averaged mean absorption
opacity, $a_{r}$ is the radiation constant, and

\begin{equation}
<\!\nu\!> = \frac{ \int_0^\infty \nu^4 n_\nu \, d\nu }
             { 4 \int_0^\infty \nu^3 n_\nu \, d\nu }
\end{equation}
is an average photon energy ($h\! <\! \nu \! >\! /k = T_r$ 
can be considered as a radiation temperature).
In the upper parts of the disk (at $z \approx H$) the density
drops to zero and since $\bar{\kappa}_{\rm ff} \propto \rho$
free--free processes are negligible in this regime. The
radiation field remains unchanged and thus $<\!\nu\!>$ and
$U$ are constant; from the boundary condition $c\, U(H) =2\, F(H)$ 
we obtain using equations (\ref{bound2}), (\ref{heightedd}), and (\ref{vturb})
for $\tau \ll 1$

\begin{equation}
\label{heatcool1}
T - T_r = \sqrt{T_v T}~,
\end{equation}
where

\begin{equation}
T_v = \left( \frac{9\alpha}{32} \right)^2
\left( \frac{\kappa_F}{\kappa_T} \right)^2 \frac{m_e^2 c^2}{\mu k m_u}
= 5.1\, 10^5\,  \alpha^2 \left( \frac{\kappa_F}{\kappa_T} \right)^2 {\rm ~K}~.
\end{equation}
(Note that in a scattering dominated plasma $\kappa_F / \kappa_T \approx 1$).
From this we get the temperature that will result from a balance
of viscous heating and inverse Compton cooling

\begin{equation}
\label{heatcool2}
T = T_r + \frac{1}{2} \left[ T_v + \sqrt{T_v^2 + 4T_r T_v} \right]~.
\end{equation}
This solution is unique since a negative sign of the square root
in (\ref{heatcool2}) would correspond to a negative sign of the
right hand side of equation (\ref{heatcool1}).
As a consequence we obtain a lower limit for $T$

\begin{equation}
T > max[T_{r},\, T_{v}]\, .
\end{equation}
\section{Results}
In this paper the mass of the black hole is fixed at a typical value of 
$M=10^{8}M_{\odot}$. 
We show results of our model calculations
for different input parameters such as the accretion rate 
$\dot{M}$, 
the viscosity parameter $\alpha$, the specific angular momentum $a$,
and the inclination angle $\Theta _{0}$ of the observer's
position
with respect to the disk axis.
Hereafter, $\dot{M}$ is measured in units of the 
critical accretion rate
$\dot{M} _{crit}=L_{Edd}/(\varepsilon c^{2})$, where
$L_{Edd}=4\pi c GM/\kappa _{T}$ is the 
Eddington luminosity and $\varepsilon $ is the efficiency of accretion.
For a non--rotating black hole ($a/M=0$) $\varepsilon$ is $0.057$, and for a
maximally rotating black hole ($a/M \simeq 0.998$, see Thorne \cite{thorne})
$\varepsilon \simeq 0.321$.     
First, we give solutions of the local vertical structure and 
emission spectrum at $5$ Schwarzschild radii 
($r=5R_{S}=10GM/c^{2}$). 
At this radius, the local  energy
release takes its maximum value for a non--rotating black hole
and this disk region 
is responsible for the high frequency part of the total spectrum.
Next, we consider the radial structure and the integrated disk spectrum
seen by a distant observer. At this point we also include 
relativistic effects on the emergent disk spectrum such as the relativistic
Doppler shift from the disk rotation, gravitational redshift, 
and gravitational light bending
due to the central black hole.
 
\subsection{Vertical Structure and Local Disk Spectrum}

\begin{figure}[htb]
\par\centerline{\psfig{figure=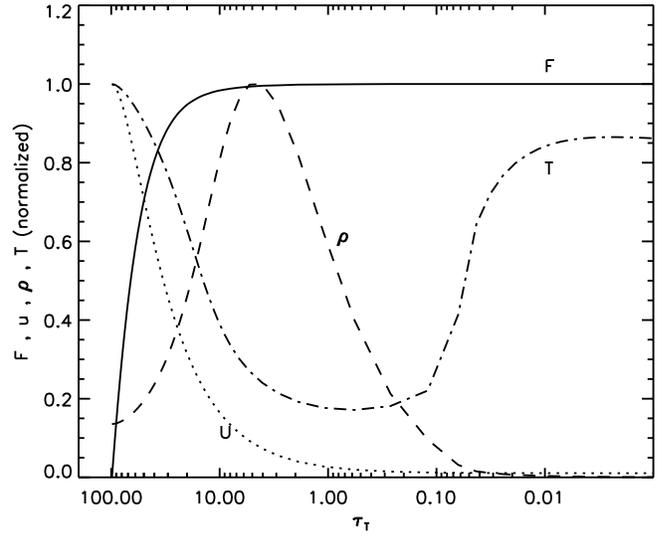,width=8.8 truecm}}
\caption{\label{vert1}Vertical distribution of radiative flux $F$, 
radiation energy density $U$,
mass density $\rho$, and gas temperature $T$ (all normalized to their
respective maximum values)
at $r=5 R_{S}$. The model parameters are: $M=10^{8}M_{\odot}$,
$\dot{M}=0.3$, $\alpha =1/3$, $a/M=0$.
Note that $U$ approaches a finite value for small optical depth
($\tau _{T} \rightarrow 0 $).}
\end{figure}

\begin{figure}[htb]
\par\centerline{\psfig{figure=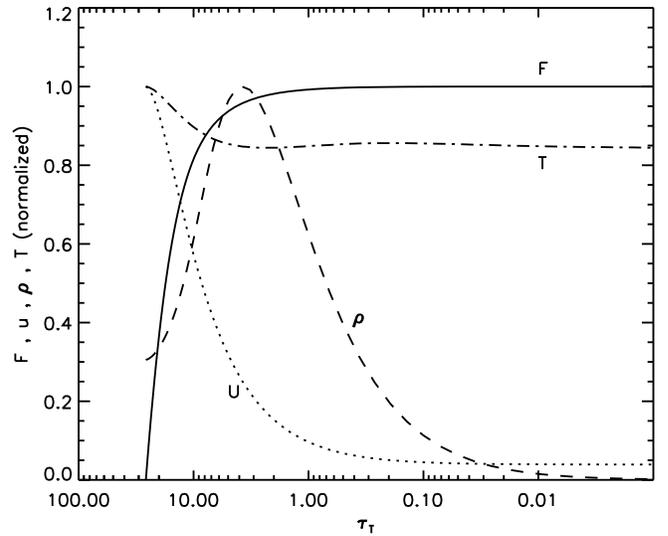,width=8.8 truecm}}
\caption{\label{vert2}Same as Fig.\ 1, but for a maximally rotating
black hole $a/M=0.998$ at $r=1 R_{S}$.}
\end{figure}

\begin{figure}[htb]
\par\centerline{\psfig{figure=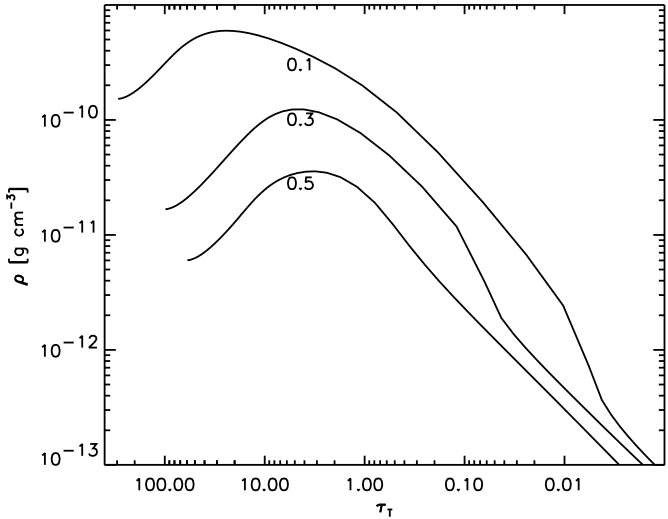,width=8.8 truecm}}
\caption{\label{dens}Vertical density distribution 
at $r=5 R_{S}$ for a model with $\alpha =1/3$ and 
$a/M=0$. The curves are labeled with the value of $\dot{M}$.}
\end{figure}

\begin{figure}[htb]
\par\centerline{\psfig{figure=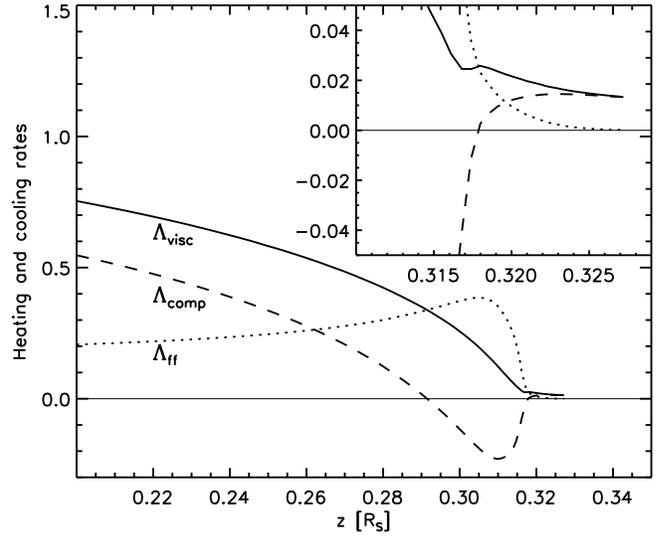,width=8.8 truecm}}
\caption{\label{heat}Viscous heating $\Lambda _{visc}$, 
Compton cooling $\Lambda _{comp}$ and
cooling due to free--free emission $\Lambda _{\rm ff}$ 
(all normalized to the viscous heating rate at the disk midplane)
as a function of $z$
at $r=5R_{S}$. The model parameters are: 
$M=10^{8}M_{\odot}$, $\dot{M}=0.3$, $\alpha =1/3$, $a/M=0$.
The outermost layer is shown in the upper right corner with 
higher resolution.}  
\end{figure}

A typical vertical distribution 
(normalized to the corresponding maximum values)
of the radiative flux $F$,
the radiation energy density $u$, the mass density $\rho $, and
the gas temperature $T$ 
as a function of the
Thomson scattering optical depth $\tau _{T}$ is shown in Fig.\ \ref{vert1} for 
a non--rotating black hole with $M=10^{8}M_{\odot}$,
$\dot{M}=0.3$, $\alpha =1/3$ and $r=5 R_{S}$. 
In Fig.\ \ref{vert2}, the above variables are plotted for the
case of
a maximally rotating black hole ($a/M=0.998$) with the same input parameters,
but using $r=1R_{S}$ , which nearly corresponds to the maximum
of the local energy release for $a/M=0.998$.
It is found that the mass density $\rho$ is not a 
monotonically decreasing function of 
$z$, but is showing a clear density inversion
(such density inversions have also been found by Meyer \& Meyer--Hofmeister
\cite{mey} and Milsom et al.\ \cite{mil}). 
This can be explained as follows:
in using the flux weighted mean $\kappa _{F}$ (equation (\ref{Rossel})),
the hydrostatic equilibrium can be written as 

\begin{equation}
\label{Hyd2}
\frac{dP_{gas}}{dz}=\frac{\kappa _{F}}{c}\rho [(F(z) - F_{Edd}(z)]
= - \rho g_{z}^{*} \, ,
\end{equation}
where the local Eddington flux $F_{Edd}(z)$
is given by 

\begin{equation}
\label{Eddi2}
F_{Edd}(z)=\frac{c}{\kappa _{F}}g_{z} \, ,
\end{equation}
and $g_{z}^{*}=g_{z}-\kappa _{F}F/c$ is the effective gravitational 
acceleration.
For a highly radiation pressure dominated disk ($P_{rad} \gg P_{gas}$), 
the local flux adjusts itself close to the Eddington value. 
For example, the radiation pressure always exceeds the gas pressure
by at least a factor of about $100$ throughout the disk for
the parameters adopted in Fig.\ \ref{vert1}. When the 
emerging surface flux is large, the local flux becomes slightly
super--Eddington in the inner regions of the disk and  
a strong density inversion ($d\rho /dz >0$)
occurs. Because of the boundary condition
imposed on the surface flux, the outer parts of the disk are
always sub--Eddington. 
In Fig.\ \ref{dens} the vertical density distribution
is shown for different $\dot{M}$. The density 
in the equatorial plane 
is up to a factor of $7$ smaller than its maximum value. 
Although the total Thomson scattering depth decreases for increasing
$\dot{M}$ and $\alpha$,
our model calculations show, that
the accretion disks are optically thick with respect to 
Thomson scattering even 
for large $\dot{M}$ and $\alpha$.    
At the same time these disks are effectively optically thin
($\tau _{\rm eff}(\nu) <1$) except for the low frequencies, 
and $\tau _{\rm eff}(\nu)$ is defined as 

\begin{equation}
\label{taueff}
\tau_{\rm eff}(\nu) =\int_{0}^{H} \rho \sqrt{3\cdot [\kappa _{\rm ff}(\nu) 
+\kappa _{T}]\kappa _{\rm ff}(\nu)} \, dz \, .
\end{equation}
In this case the assumption of local thermodynamic equilibrium (LTE) 
is no longer valid even in the disk midplane and  
the gas temperature deviates strongly from the equilibrium
temperature defined as $T_{eq}=(U/a_{r})^{1/4}$.
The various heating and cooling functions are compared in 
Fig.\ \ref{heat}. 
Compton cooling plays the dominat role in the
inner layers of the disk (at small $z$), whereas in the region of the
density increase free-free-processes become very efficient leading to
a steep temperature drop. The gas temperature even falls below the
radiation temperature $T_r$, thus Compton scattering turns into a
heating mechanism ($\Lambda_{comp}$ is negative) which effectively balances the
bremsstrahlung cooling. In the uppermost layers, where the density drops
exponentially, free-free processes are negligible ($\propto \rho^2$),
and the temperature adjusts itself to the equilibrium between viscous
heating and inverse Compton cooling described above.
Due to these heating-cooling processes a temperature inversion occurs
and a hot corona-like layer builds up on top of the disk. This layer
is however optically thin to both scattering and 
absorption (the typical densities are below
$10^{-13}\, {\rm g\, cm^{-3}}$), and thus there is no effect on the emitted
spectrum.
Note that in the upper disk layers the viscous heating is proportional
to the density, thus the relevant cooling mechanisms must not 
have a stronger $\rho$--dependence. This holds of course 
for inverse Compton cooling used in our model, however, 
cooling due to strong resonance lines has also been proposed 
(Hubeny \cite{hubeny}) to contribute to the overall heating--cooling 
balance.
 
\begin{figure}[htb]
\par\centerline{\psfig{figure=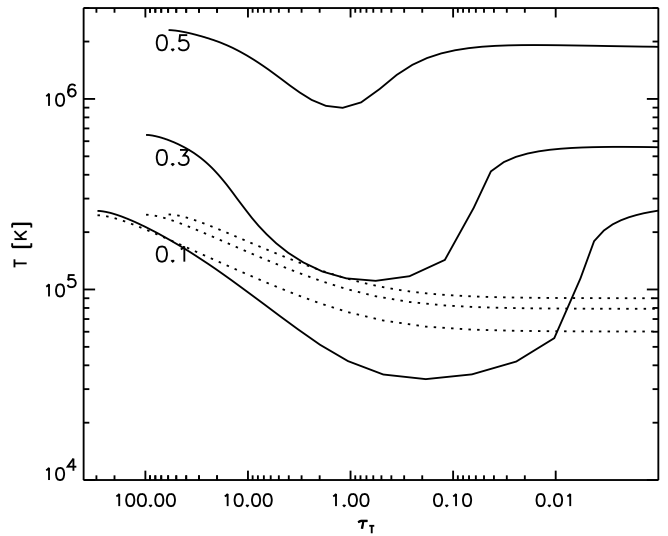,width=8.8 truecm}}
\caption{\label{temp1}Vertical distribution of gas temperature $T$
(solid lines)
and equilibrium temperature $T_{eq}$ (dotted lines)
at $r=5R_{S}$ for a model with $\alpha =1/3$ and 
$a/M=0$. The curves are labeled with the value of $\dot{M}$.}
\end{figure}

The gas temperature and the corresponding
equilibrium temperature $T_{eq}$ for
different $\dot{M}$ are shown in Fig.\ \ref{temp1}.

For large values of $\dot{M}$ even the temperature
in the midplane of the disk strongly exceeds the equilibrium value.
For low  $\dot{M}$ the disk becomes effectively optically
thick and a general equilibrium is established.

\begin{figure}[htb]
\par\centerline{\psfig{figure=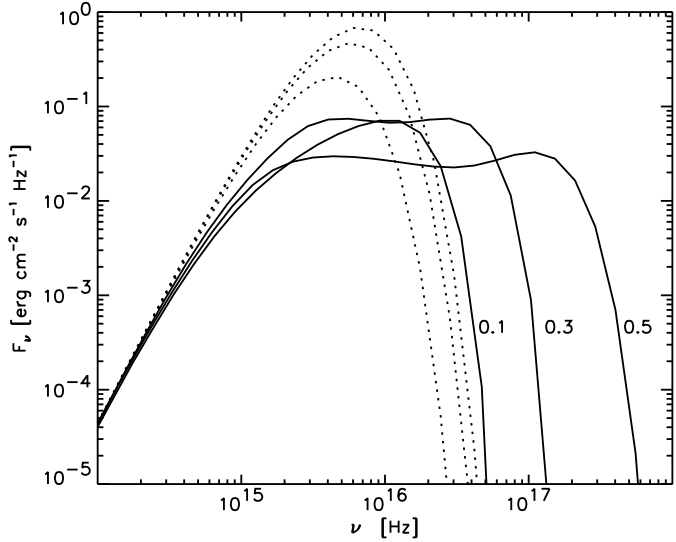,width=8.8 truecm}}
\caption{\label{spec1}Emergent spectrum 
at $r=5R_{S}$ for $\alpha =1/3$ and $a/M=0$.
The curves are labeled with the value of $\dot{M}$.
Dotted curves show the 
local Wien spectrum at $T=T_{\rm eff}$.}
\end{figure}

The locally emitted spectrum at a fixed radial position 
($r=5R_{S}$) for 
different $\dot{M}$ is shown in 
Fig.\ \ref{spec1}.
The local Wien spectra at the corresponding effective temperature $T_{\rm eff}$
are also shown by dotted lines. At low frequencies, the accretion disk is 
effectively optically thick and scattering effects can be neglected;
the local spectrum approaches the equilibrium distribution $W_{\nu}$,
however, the spectra in low frequency regime cannot be taken too seriously
since we have neglected induced processes.
At higher frequencies,
scattering effects become important and one gets a modified Wien spectrum.
For sufficiently high $\dot{M}$ and $\alpha$ a Wien peak appears 
in the soft X--ray range of the local spectra. This feature is a result
of repeated Compton scattering of photons by a
thermal distribution of electrons at temperature $T$. If 
the Compton parameter $y \gg 1$,
Comptonization goes to saturation and the photons are shifted 
into a Wien distribution at temperature $T$ 
(see Felten \& Rees \cite{felten}).
The Compton parameter $y$ is defined as  

\begin{equation}
\label{comp}
y=\frac{4kT}{m_{e}c^{2}}\tau _{T}^{2} \, .
\end{equation}
For example, the case
$\dot{M}=0.3$ shown in Fig.\ \ref{spec1}
leads to a effective optical depth  
$\tau _{\rm eff}(\nu _{max})=0.1$ at the peak frequency
$\nu _{max}=2.8 \cdot 10^{16}\, Hz$ 
and a total Thomson
scattering depth of $\tau _{T}=98.5$;
from that and an average temperature of
$T=4\cdot 10^{5} K$ (see Fig.\ \ref{temp1}) we have 
$y \approx 3$.
Since repeated
Compton scattering is important for $y >1$, a significant shift
of photons
into the Wien peak is expected.
For low $\dot{M}$ and $\alpha$ the disk is effectively optically
thick for all relevant frequencies
and the radiation assumes an equilibrium spectrum for almost all $z$.
The effective optical depth 
for the curve $\dot{M}=0.1$
in Fig.\ \ref{temp1} and \ref{spec1} is $\tau _{\rm eff}(\nu _{max})=9.8$,
with $\nu _{max}=9.1 \cdot 10^{15}\, Hz$.
Nevertheless, the emergent spectrum deviates from the Wien spectrum
at $T=T_{\rm eff}$ because there exists a temperature gradient and 
high frequency photons escape from deeper layers with higher
temperatures.

\subsection{Radial Structure and Integrated Disk Spectrum} 

\begin{figure}[htb]
\par\centerline{\psfig{figure=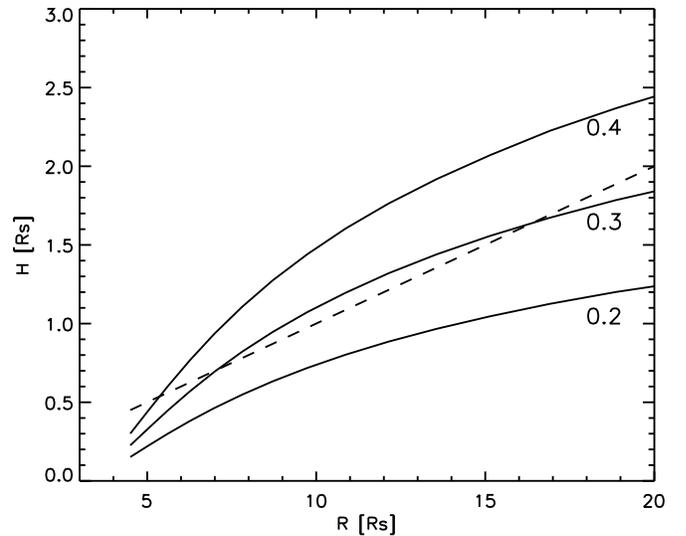,width=8.8 truecm}}
\caption{\label{height1}Disk height
for $\alpha=1/3$ and $a/M=0$.
The curves are labeled with the value of $\dot{M}$.
The dashed curve shows the geometrically thin limit
$H(r)=0.1 \cdot r$.}
\end{figure}

After discussion of the vertical structure and local 
emission spectrum at $r=5R_{S}$ here
we investigate the radial structure and overall spectrum
of the accretion disk. In numerical calculations,
we used $50$ radial points on a logarithmic grid 
from the last stable orbit $r_{i}$ to the outer disk radius 
$r_{out}$ (here $r_{out}=1000R_{S}$).
The height of the disk from the surface to the
equatorial plane as a function of the radial distance
from the black hole is shown in Fig.\ \ref{height1} 
for several values of $\dot{M}$.
For comparison, the geometrically thin limit, here defined as
$H(r)=0.1 r$, is also plotted as a dashed curve.
This means that the accretion rate should not 
greatly exceed $\dot{M}=0.3$
for not to violate the geometrically thin disk approximation.
Therefore, numerical results for higher accretion rates are 
strictly speaking not self--consistent.
On the other hand, the $\alpha$--parameter has little influence 
(of the order of a few per cent) on the
height of the disk.
If the local flux is given by the Eddington flux,
the disk height $H$ is given by equation (\ref{heightedd}). 
In the inner regions of the disk Thomson scattering delivers the major
contribution to the opacity ($\kappa _{F} \approx \kappa _{T}$)
and the height of the disk is solely determined by the
accretion rate $\dot{M}$.
However, the local flux deviates from the Eddington flux at the upper
boundary of the disk. Although the gas pressure is negligible in comparison
with the radiation pressure, the gas pressure gradient $dp_{Gas}/dz$
is not.
We have also determined mean radial velocities 
$|v_{r}| = \dot{M}/ \Sigma (r) $ for all radial grid points.
Our calculations show, that for $\dot{M} \leq 0.3$ and
$\alpha \leq 1$ the basic thin disk approximation   
$|v_{r}| \ll |v_{\phi}| $ is always satisfied.
\begin{figure}[htb]
\par\centerline{\psfig{figure=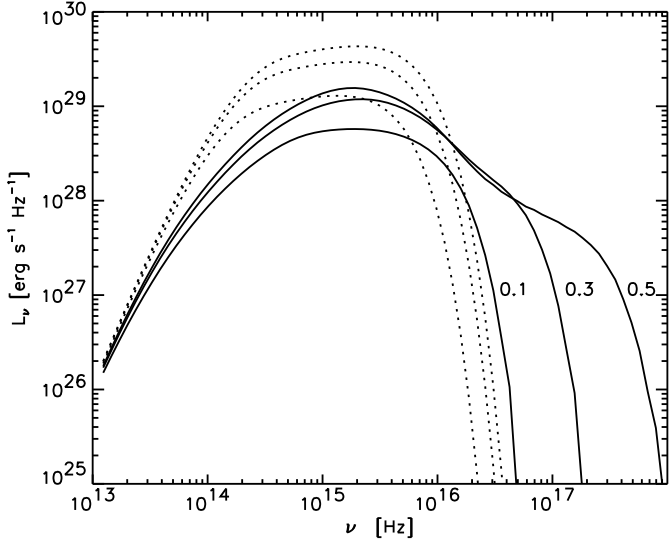,width=8.8 truecm}}
\caption{\label{newt1}Entire Newtonian face on disk spectra 
for $\alpha=1/3$ and $a/M=0$.
The curves are labeled with the value of $\dot{M}$.
Dotted curves denote multi Wien spectra.}
\end{figure}
\begin{figure}[htb]
\par\centerline{\psfig{figure=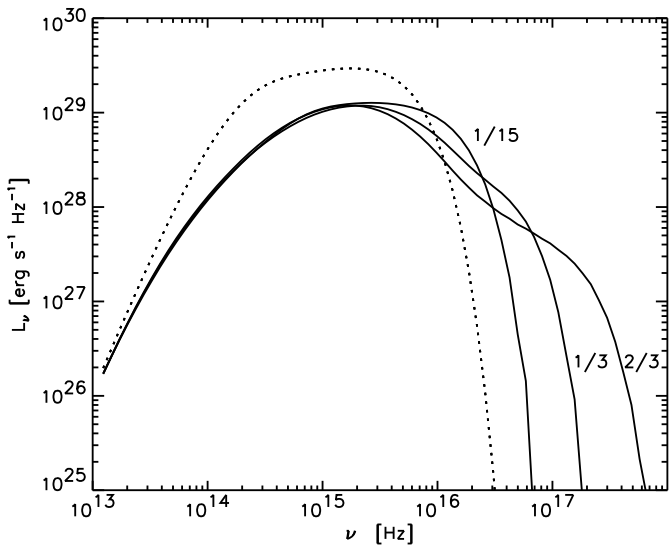,width=8.8 truecm}}
\caption{\label{newt2}Entire Newtonian face on disk spectra
for $\dot{M}=0.3$ and $a/M=0$.
The curves are labeled with the value of $\alpha$.
The dotted curve denotes the corresponding multi Wien spectrum.}
\end{figure}     
Now we will turn to the integrated disk spectra.
General relativistic effects on the emergent spectrum are twofold:
first, relativistic effects occur in calculating the 
disk structure and therefore change the local emission spectrum in
the corotating frame of the disk. The resulting spectra were
discussed for a non--rotating black hole in section 3.1.
Second, relativistic effects have a substantial influence on the 
propagation of photons to the observer. All these effects become
more and more important if the black hole is rotating, since then 
the disk extends to regions very near to the black hole.
To separate both effects the investigation of the integrated disk spectrum 
is treated in
two different ways:
first, we examine the emergent spectrum of a face on disk as
seen for a distant observer without including relativistic effects
or Doppler shifts
on the locally emitted photon distribution. Relativistic effects only occur
due to the relativistic disk structure equations (see section 2.1).
The entire spectrum then is simply calculated by integration
of  
the local spectra over the disk surface. In the following we will
call this the Newtonian approximation.
Second, a fully relativistic calculation
of the propagation of photons from the disk to a distant observer
is performed, using a program code (Speith et al.\ \cite{speith}) 
to obtain numerical values of the Cunningham transfer function
(Cunningham \cite{cunningham})
for any set of parameters.

\begin{figure}[htb]
\par\centerline{\psfig{figure=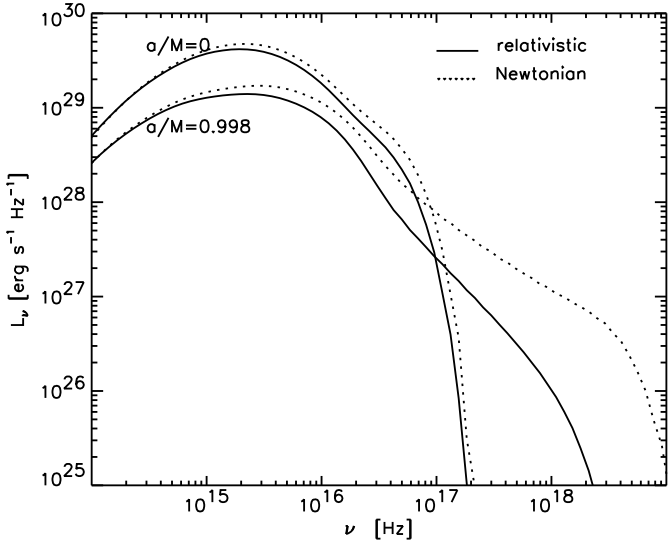,width=8.8 truecm}}
\caption{\label{face}Entire face on disk spectra for 
$\dot{M}=0.3$ and $\alpha=1/3$. 
The curves are labeled with the value of the specific angular momentum $a/M$.
The solid an dotted curves denote the relativistic and 
Newtonian disk spectra, respectively.}
\end{figure}

\begin{figure}[htb]
\par\centerline{\psfig{figure=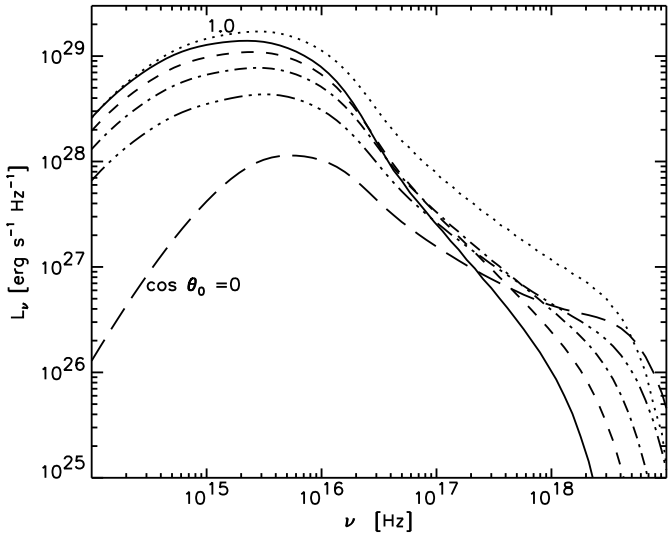,width=8.8 truecm}}
\caption{\label{theta}Entire relativistic disk spectra
for different inclination angle $\Theta _{0}=0\degr ,\,
41\degr ,\, 60\degr ,\, 70\degr $ and $90\degr$. The model parameters are:
$M=10^{8}M_{\odot}$, $\dot{M}=0.3$, $\alpha=1/3$, $a/M=0.998$.
The dotted curve denotes the Newtonian face--on spectrum.}
\end{figure}

The Newtonian disk spectrum for several
values of $\dot{M}$ and $\alpha$ is shown in Fig.\ \ref{newt1}
and \ref{newt2}. The 
spectral luminosities calculated by summing up local Wien spectra at
the corresponding effective temperatures 
are indicated by dotted lines. The hardness of the spectra and 
therefore also the amount of soft X--rays is a sensitive 
function of $\dot{M}$ and a somewhat less sensitive 
function also of $\alpha$.
In the following, we fix the accretion rate and viscosity parameter
at $\dot{M}=0.3$ and $\alpha =1/3$, respectively. 
A comparison of the relativistic and Newtonian face--on disk spectrum
for a non--rotating ($a/M=0$) and maximally rotating ($a/M=0.998$) black hole is
shown in Fig.\ \ref{face}. 
For low frequencies the emergent spectrum approaches the Newtonian
model. The spectrum at higher frequencies is reduced 
with respect to the Newtonian case especially for a rotating black hole.
These photons originate from 
the inner parts of the disk with high velocities and
gravitational fields and therefore are diverted by forward peaking and 
gravitational focusing. 
Consequently, an equatorial observer primarily sees blueshifted and
focused radiation from the hot inner parts of the disk.
In Fig.\ \ref{theta} the observed relativistic spectrum is shown as a 
function of the
inclination angle $\Theta _{0}$ together with a 
Newtonian face--on spectrum (dotted line) for a maximally rotating black hole.
A significant fraction of total flux is emitted in the soft and hard 
X--ray range 
($\nu > 2.4\cdot 10^{16}Hz$) and strongly depends 
on the position of the observer. For 
$\Theta _{0} = 0\degr ,\, 41\degr ,\, 60\degr ,\, 70\degr $ 
and $90\degr$,
this fraction is $36\%$, $50\%$, $64\%$, $78\%$
and $93\%$, respectively. The fraction of the total flux emitted in the 
soft X--ray band ($2.4\cdot 10^{16}\, Hz-6\cdot 10^{17}\, Hz$), 
corresponding to the 
sensitivity range of the PSPC (position sensitive proportional
counter) on the ROSAT X--ray satellite is $32\%$, $38\%$, $38\%$, $33\%$
and $21\%$, respectively, and therefore is nearly independent of the 
inclination angle.
In order to show the limiting case of our calculations we also
included $\cos \Theta _{0} = 0 $ in Fig.\ \ref{theta}.
However, for a nearly edge-on observer self--occultation of the
inner parts of the disk by the outer parts should be taken into account. 

\section{Concluding Remarks}

We have performed a self--consistent calculation of the 
vertical structure and emergent spectrum of an accretion disk
around a massive Kerr black hole.
Full relativistic corrections have been
included.
The standard $\alpha$--model leads to diverging temperature profiles
in the upper layers of the disk, where the density drops exponentially
and viscous heating always overcomes radiative cooling.
When we include the radiative cooling of the turbulence elements in the
optically thin part of the disk, an equilibrium between turbulent 
viscous heating and inverse Compton cooling is established.
Our calculations are for a fixed central mass $M=10^{8}M_{\odot}$
but for different $\dot{M}$, $\alpha$, $a/M$, and $\Theta _{0}$.
It is found that the mass density is not a monotonically decreasing
function of $z$ but shows a strong density inversion in the
radiation pressure dominated inner parts of the disk, where
the local flux becomes slightly super--Eddington. Moreover, we always obtain
a temperature inversion in the upper optically thin layers of the disk
with zero gradient in the outermost region.
For sufficiently high $\dot{M}$ and $\alpha$ the disk becomes 
effectively optically thin at small radii and the gas 
temperature exceeds the equilibrium temperature 
($T_{eq}=(U/a_{r})^{1/4}$) by a large amount even in the disk midplane.
As a result, the local emission spectrum strongly deviates from the
Wien spectrum at $T=T_{\rm eff}$.
Even in the optically thick case, the local spectra differ from the 
Wien--spectra at high frequencies, because scattering opacity dominates 
in the soft X--ray range and there exists a temperature gradient
in the vertical direction. Therefore, a significant fraction 
of the total flux is emitted in the soft X--ray range even for low 
$\dot{M}$
and $\alpha$, especially in the maximally rotating case ($a=0.998$)
and for high inclination angles of the observer.
For increasing $\dot{M}$, $\alpha$, $a/M$, and $\Theta _{0}$
the calculated spectra become more and more flat, producing quasi power law
spectra 
in the sensitivity range of the ROSAT PSPC, whereas for small values 
a steep soft X-ray component is established.
Thus, the model can in general account for the soft X--ray excess 
observed in many AGN. 
We intend to show detailed spectral fits of this model to combined ROSAT and 
IUE observations for a sample of AGN in a future paper. 
   
Note that we have neglected induced processes and that our opacity description
only contains free--free processes for a pure hydrogen atmosphere.
Neglecting induced processes means, that we have concentrated 
on the high energy tail of the resulting spectra.
Nevertheless, bound--free absorption and line opacities can significantly
contribute to the total opacity especially at low temperatures.
Therefore, the spectrum is expected to soften when all relevant opacities
are taken into account because of the increasing effective optical depth
in this case. 

Another important point is the role of convection in our calculations,
especially in view of our inverse density profiles.
If convective energy transport takes place in the disk, 
the vertical structure and subsequently the emergent spectrum may be altered
considerably.
We therefore have done a simplified stability analysis according to 
the standard mixing length theory (see for example Cox \& Giuli \cite{cox}),
assuming that the convective elements move adiabatically in the disk
and that the local gas temperature is given by the equilibrium temperature
$T_{eq}$. In a gas with radiation pressure, the effective buoyancy force
is reduced by the radiation pressure gradient and the convective flux 
then can be expressed in terms of the effective gravitational acceleration
$g_{z}^{*}$ (equation \ref{Hyd2}),
\begin{equation}
\label{convec}
F_{conv}=\frac{\rho c_{P}Tl_{m}^{2}H_{P}^{-3/2}\delta ^{1/2}}{4\sqrt{2}}
\sqrt{g_{z}^{*}(\nabla _{s} -\nabla _{e})}\cdot (\nabla _{s} -\nabla _{e})\, ,
\end{equation}    
where $l_{m}$, $H_{P}=P/(\rho g_{z})$, and $c_{P}$ are the mixing length,
the pressure scale height and the specific heat at constant pressure,
respectively, and $\delta=-\partial ln\rho /\partial lnT$.
$\nabla _{s}=(dlnT/dlnP)_{s}$ and $\nabla _{e}=(dlnT/dlnP)_{e}$
are the temperature gradients with respect to pressure
for the surroundings and the rising element.
Therefore, the disk is unstable for convection if the square root 
on the right hand side of equation (\ref{convec}) has a positive argument.
In the sub--Eddington regime one has $g_{z}^{*}>0$ and
the condition for convection to set in
is given by the Schwarzschild criterion $(\nabla _{s} -\nabla _{e})>0$.
On the other hand in the super--Eddington regime $g_{z}^{*}<0$,
and the condition for convective instability is
$(\nabla _{s} -\nabla _{e})<0$, leading to a convective energy flux in 
negative $z$--direction.
Our model calculations show that none of both conditions are satisfied
and the disk therefore is stable against convection.
Since the gas temperature deviates from the equilibrium temperature for
an effectively optically thin disk and the convective elements 
do not move adiabatically in the upper optically thin layers, this
simplified stability analysis 
has to be considered with caution, and a more detailed
stability analysis has to be performed in order
to clarify the role of convection.

It is well known that an $\alpha$-disk with a viscosity proportional
to the total pressure is radially unstable in the radiation
dominated case (Lightman \& Eardly \cite{light}).
However, this analysis is based on a vertically integrated disk
structure. Since our model leads to a strong $z$-dependence for almost
all functions including the viscosity, a detailed 2-dimensional
stability analysis would be required for a definite answer
concerning the overall stability of our model.

\begin{acknowledgements}
T.\ D\"orrer acknowledges the support of DARA through grant 50 OR 90099.
\end{acknowledgements}

{}


\begin{thebibliography}{}
\bibitem[1985]{arnaud}
Arnaud, K.\ A.\ , Branduardi--Raymont, G.\ , Culhane, J.\ L.\ , 
et al.\ , 1985, MNRAS, 217, 105
\bibitem[1987]{bechtold}
Bechtold, J.\ , Czerny, B.\ , Elvis, M.\ , Fabiano, G.\ , Green, R.\ F.\ , 
1987, ApJ, 314, 699
\bibitem[1988]{carson}
Carson, T.\ R.\ , 1988, A\& A, 189, 319
\bibitem[1968]{cox}
Cox, J.\ P.\ , \& Giuli, R.\ T.\ , 1968, in 
{\it Principle of Stellar Structure},
Volume 1 (Physical Principles), Gordon \& Breach, New York
\bibitem[1975]{cunningham}
Cunningham, C.\ T.\ , 1975, ApJ, 202, 788
\bibitem[1987]{czerny}
Czerny, C.\ T.\ , \& Elvis, M.\ , 1987, ApJ, 321, 305
\bibitem[1972]{felten}
Felten, J.\ E.\ , \& Rees, M.\ J.\ , 1972, A\& A, 17, 226
\bibitem[1990]{hubeny}
Hubeny, I.\ , 1990, ApJ, 351, 632
\bibitem[1961]{karzas}
Karzas, W.\ J.\ , \& Latter, R.\ , 1961, ApJS, 6, 167
\bibitem[1957]{kompanaets}
Kompanaets, A.\ S.\ , 1957, Soviet Phys.\ JETP, 4, 730
\bibitem[1989]{laor89}
Laor, A.\ , \& Netzer, H.\ , 1989, MNRAS, 238, 897
\bibitem[1990]{laor90}
Laor, A.\ , Netzer, H.\ , Piran, T.\ , 1990, MNRAS, 242, 560
\bibitem[1974]{light}
Lightman, A.\ P.\ , \& Eardley, D.\ M.\ , 1974, ApJ, 187, L1
\bibitem[1982]{malkan}
Malkan, M.\ A.\ , \& Sargent, W.\ L.\ W.\ , 1982, ApJ, 254, 22
\bibitem[1982]{mey}
Meyer, F.\ , \& Meyer--Hofmeister, E.\ , 1982, A\& A, 106, 34
\bibitem[1994]{mil}
Milsom, J.\ A.\ , Chen, X.\ , Taam, R.\ E.\ , 1994, ApJ, 421, 668
\bibitem[1973]{novikov}
Novikov, I.\ D.\ , \& Thorne, K.\ S.\ , 1973, in {\it Black Holes},
eds.\ de Witt, C.\ \& de Witt, B.\ , Gordon \& Breach, New York
\bibitem[1974]{page}
Page, D.\ N.\ , \& Thorne, K.\ S.\ , 1974, ApJ, 191, 499
\bibitem[1990]{pounds}
Pounds, K.\ A.\ , Nandra, K.\ , Stewart, G.\ C.\ , 
George, I.\ M.\ , Fabian, A.\ C.\ , 1990, Nat, 344, 132
\bibitem[1972]{pringle}
Pringle, J.\ E.\ , \& Rees, M.\ J.\ , 1972, A\& A, 21,1
\bibitem[1995]{riffert}
Riffert, H.\ , \&  Herold, H.\ , 1995, ApJ, in press
\bibitem[1992]{ross}
Ross, R.\ R.\ , Fabian, A.\ C.\ , Mineshige, S.\ , 1992, MNRAS, 258, 189
\bibitem[1973]{shakura}
Shakura, N.\ I.\ , \& Sunyaev, R.\ A.\ , 1973, A\& A, 24, 337
\bibitem[1978]{shields}
Shields, G.\ A.\ , 1978, Nat, 272, 706
\bibitem[1993]{shimura}
Shimura, T.\ , \& Takahara, F.\ , 1993, ApJ, 419, 78
\bibitem[1995]{shimura2}
Shimura, T.\ , \& Takahara, F.\ , 1995, ApJ, 440, 610
\bibitem[1995]{speith}
Speith, R.\ , Riffert, H.\ , Ruder, H.\ , 1995, Comp.\ Phys.\ Comm.\ , 
in press
\bibitem[1974]{thorne}
Thorne, K.\ S.\ , 1974, ApJ, 191, 507
\bibitem[1988]{wandel}
Wandel, A.\ , \& Petrosian, V.\ , 1988, ApJ, 329, L11
\bibitem[1987]{wilkes}
Wilkes, B.\ J.\ , \& Elvis, M.\ , 1987, ApJ, 323, 243 
\bibitem[1989]{turner}
Turner, T.\ J.\ , \& Pounds, K.\ A.\ , 1989, MNRAS, 240, 833
\bibitem[1994]{yamada}
Yamada, T.\ T.\ , Mineshige, S.\ , Ross, R.\ R.\ , Fukue, J.\ , 1994, 
PASJ, 46, 553
\end{thebibliography}
\end{document}